# Low-Cost Architecture for an Advanced Smart Shower System Using Internet of Things Platform

Shadeeb Hossain[1†] [ORCID ID: 0000-0002-5224-7684], Ahmed Abdelgawad[1]

[1] School of Engineering and Technology , Central Michigan University, MI-48859

***Abstract***

Wastage of water is a critical issue amongst the various global crises. This paper proposes an architecture model for a low-cost, energy efficient "SMART Shower" system that is ideal for efficient water management and be able to predict reliably any accidental fall in the shower space. The sensors in this prototype can document the surrounding temperature and humidity in real time and thereby circulate the ideal temperature of water for its patron, rather than its reliance on predictive values . Three different scenarios are discussed that can allow reliably predicting any accidental fall in the shower vicinity. Motion sensors, sound sensors and gesture sensors can be used to compliment prediction of possible injuries in the shower. The integration with the Internet of Things (IoT) platform will allow caretakers to monitor the activities in the shower space especially in the case of elderly individuals as there have been reported cases of casualties in the slippery shower space. The proposed proof-of-concept prototype is cost effective and can be incorporated into an existing system for the added precedence of safety and convenience. The intelligent system is conserving water by optimizing its flow temperature and the IoT platform allows real time monitoring for safety.

**[‡]Corresponding Author: shadeebhossain@yahoo.com**

## I. INTRODUCTION

Individuals across the globe are making their own contributions towards conserving water and energy but an average US household waste annually 1000s of gallons of water in miscellaneous applications [1-4]. The concept of smart cities promotes sustainability in terms of both energy and waste management [5,6]. The Internet of Things (IoT) platform can assist in achieving the desired goal of efficient management with the application of sensors, software programming and exchange of relevant data over the cloud. It is predicted that in the future approximately 3 in every 10 urban buildings will utilize IoT technology or operate on smart devices [7,8].

The integration of Wireless Sensor Network (WSN) with IoT platform has provided opportunities for different domestic applications [9-11]. The architecture for a low power, Cloud based IoT application [12, 13] can be easily employed in SMART homes to monitor various conditions including temperature, humidity, sound, light intensity, etc.

The bathroom or the shower space is an integral component of an individual's house, and it is associated with everyday activities. The traditional shower facility has evolved over the years and has been accommodated for various functions. Hirsch *et al.* (1998) introduced the concept of smart shower with a patented design that regulates its operational flow via monitoring proximity range [14]. The primary goal was to introduce an improved experience for the patrons which allows an economical, contactless operation without any complex mechanical actuation system. Rodrigues *et al.* (2018) also developed an IoT based smart electric shower that analyzed various parameters including water and power consumption, water flux and bath temperature [15]. The goal was to analyze the utilization pattern and predict savings on the utility bill. Bodhe *et al.* (2016) developed a smart shower system called Salle de B(r)ains that used WSN and integrated IoT to gauge water and power consumption [16]. It used Wi-Fi communication and cloud data analytics to mitigate water wastage and to encourage users to realize their water consumption. Lucas (2018) also worked on developing a prototype for "*Enviro Shower*" system, where efforts were made to regulate the flow rate and *"times shower valve manager"* that included a shower timer and flow control device in the system design [17].

The improving life expectancy of the growing population and to assist individuals with disabilities, a smart shower or water management system can allow independence in navigation or even call an immediate healthcare provider due to an unfortunate mishap in the bathroom vicinity. Ma *et al.* (2017) proposed an adaptive assistive smart shower system capable of detecting disability of the patron and providing the necessary aids automatically [18]. Gesture based interfaces have also been proposed by Chen *et al. (2022)* to perform interactive tasks in the shower room [19]. Their study however only focused on showing that 71.6% of participants preferred foot gestures as opposed to hand gestures. They also proposed guidelines for foot gesture interfaces based on models and behaviors while interacting with foot gestures. Ferati *et al.* (2018) also co-designed a smart shower for people with disabilities [20]. They used an electronic building block called *littleBits* that included features such as: (i) turning on the device (ii) simulation of the shower on-off (iii) temperature monitoring.

The smart shower therefore should address the challenge of (i) water conservation and (ii) shower space causality and injuries. An attempt has been made in this proposed model to expatiate the issues relating to (1) *consumer convenience* (2) *environmentally friendly* and (3) *safety and reliance*. Apart from consumer convenience, the market should also target consumer safety in vulnerable shower confinement by accurately predicting an individual's accidental fall in the confined shower space.

Our proposed prototype has the evolving consumer needs for the design of a smart shower system. The proximity measurement via waterproof ultrasonic sensor ensures moderate water consumption. The sensor can also be integrated with the microcontroller to allow real-time tracking via IoT for any accidental fall in the slick shower space. According to the census of patients admitted to Stoke Mandeville Hospital (Aylesbury, UK), it was realized that burns and scalds due to hot water can be lethal and reported for significant number of casualties [21]. The temperature and humidity sensor can therefore allow real-time tracking of external temperature and be programmed to regulate the 'preferred' antithetical water flow temperature. The microcontroller can be programmed to accommodate for individual patron user's preferences that shares the common household shower booth (for instance, *user A* prefers cold as opposed to *user B* who always prefers the lukewarm water temperature); it can account for additional safety by ensuring maximum water threshold to reach is below $50^0$C to avoid scalds. The proposed system can be easily implemented into an existing shower system and is relatively cost effective.

The purpose of this prototype is to develop a proof-of-concept cost-effective smart shower system using Raspberry Pi 3 that primarily helps in (i) *efficient water management* and (ii) *monitor shower space "accidents"*. The IoT platform for this application is Thing Speak IoT and the sensors include (i) HC-SR04 ultrasonic sensor (ii) AH10-temperature and humidity sensor (iii) CZN -15E sound sensor and (iv) PAJ 7620 gesture sensor. Python programming code was used to interface the sensors with Raspberry Pi 3.

## II. FRONT END ARCHITECTURE FOR THE PROPOSED SMART SHOWER SYSTEM

The proposed architecture of the consumer end of the shower space will provide for advanced features as shown through Fig. 1. The strategic placement of the waterproof ultrasonic sensor at a measured height of approximately 2 feet can ensure proximity measurement for its controlled activation and deactivation via a pre-programmed microcontroller. The shower head will only activate when an individual is within proximity. This feature will This installed feature is environmentally friendly as it accounts for moderate consumption. HC-SR04 ultrasonic sensor was used for this proof-of-concept but waterproof JSN-SR04T-2.0 ultrasonic sensor is a suitable alternative in the actual application.

**Fig. 1 HERE**

A sound sensor can also be integrated into the system as it could allow detection of a sudden fall in the shower space or a desperate cry for help. CZN -15E microphone-based sound sensor can be used in this application. The output of the sensor is binary and is correlated to the intensity of the sound, for instance, the output signal is binary 1 if the sound intensity is above a particular threshold. This could in turn trigger a call to the emergency or a responsible guardian that requires immediate attention.

A gesture sensor such as PAJ7620 will allow the flexibility of using motion or gestures when an individual is having difficulty to call out for help (due to falling in the shower space) or too occupied to access features in the shower space (such as motioning to turn off shower or switch to a cooler water). PAJ7620 can recognize up to nine gestures such as moving up, down, right, left, forward, backward, clockwise, anticlockwise and waving. Each gesture can be programed to execute a particular task.

The temperature and humidity sensor such as DHT 11, can be placed outside the building to monitor the fluctuating temperature. It can account for real time data measurement to ensure antithetical water discharge through the shower head (for instance if it is a hot sunny day, it would discharge a cold shower water and vice versa). This would aviate the time for the shower water to culvert to the desired temperature. The parallel operation programming with the ultrasonic sensor will avoid power consumption and any undesired water discharge.

The sensors can be connected wirelessly through Bluetooth or through a wired network to the microcontroller. The microcontroller can be further connected to a keypad installed directly outside the shower confinement. This will allow an added flexibility to individual houseowners to preset their preferred shower temperature. The shower water temperature can however have thresholds at 50 $^{0}$C to avoid scalds and burn injuries.

To account for safety and reliance, the IoT platform is unparalleled. The real time monitoring of shower space occupancy can also be directly correlated to accidental injuries. A signal can be sent to the responsible guardian or caretaker if the occupancy exceeds a certain time threshold. Additional ultrasonic proximity sensors can be placed in shower space confinement for a more accurate measurement and the microcontroller can analyze those data prior to sending the signal.

The gathered data can be analyzed through a customized app for water consumption, power consumption, flow rate, the temperature/humidity fluctuation. These primary data are critical to monitor water management and can be used in analytics to keep track of water and energy wastage. The PAJ7620-gesture sensor and CZN-15E sound sensor can complement monitoring the duration spent by an elderly or child in the shower space and in turn send a peril implied signal if an abnormality is detected. Everyone sharing a common shower space can possess their individual 4- digit code that could be entered through keypad prior to entering the shower space. Fig. 2 shows a proposed graphical mobile user interface at the consumer end displaying information about his shower time data.

**Fig. 2 HERE**

## III. ARCHITECTURE FOR SMART SHOWER USING IOT

The architecture for the prototype consists of three layers: (1) *Hardware Level* (2) *Cloud Server* (3) Algorithm *Development*.

### A. *Hardware Level*

The hardware for the proposed architecture consists of *(a) Microprocessor (b) Wireless Sensors (c) Actuators (d) Wireless Communication Module* as shown in Fig. 3.

**Fig.3 HERE**

Raspberry Pi 3 with Quad-core 64-bit ARM Cortex A53 is used as the microprocessor during the development of the project. Raspberry Pi 3 has the advantage of being a small portable microprocessor and has two 5V pins, two 3.3V pins and nine ground pins. Raspberry Pi has the feature of 21 GPIO (General Purpose Input/Output) pins for connecting the sensors/actuator, USB ports and has low power consumption. Raspberry Pi and Arduino have been used in the design of multiple smart home applications. Vujovic *et al*. (2015) used Raspberry Pi as a sensor web node for the design of a home automation system with an application of monitoring and determination of the confidence of fire in a building [22]. Ferdoush *et al.* (2014) similarly used Raspberry Pi and Arduino for environmental monitoring applications [23]. Noor (2013) also developed a cost effective and scalable community-based home security system using Raspberry Pi that was interfaced with the sensor to detect movement [24].

For this prototype bench testing, wired ethernet connection was used to connect the Raspberry Pi to the network router using a RJ45 cable. However, wireless Wi-Fi connection is already built in the Raspberry Pi 3 Model B used in this initial prototype. The built-in Wi-Fi module was used to integrate with the Cloud Server and hence allowed taking the advantage of its integration with IoT. Raspberry Pi Operating System (OS) has a user-friendly interface that helps to connect the relevant SSID – (Wi-Fi network). However, in the case of a scalable and marketable prototype, the headless Raspberry Pi can be connected to the IoT platform using the following steps shown in Fig.4.

**Fig. 4 HERE**

The sensors and the actuators were all wired to the input/output pins of the Raspberry Pi 3 as shown earlier in Fig. 3. I2C (inter-integrated controller) serial communication was used for transferring data between Raspberry Pi 3 and the sensors/actuators. The following steps were followed to establish the I2C communication in the Raspberry Pi.

| | |
|---|---|
| | sudo raspi-config |
| *Select* | 5-Interfacing Options |
| *To check devices connected* | I2Cdetect -y 1 |
| *To manage I2C communication* | pip3 install smbus2 |

Waterproof JSN-SR04T-2.0 ultrasonic sensor would be ideal for proximity measurement for this example. Its range varies between 20 cm to 600 cm and can be easily integrated with Raspberry Pi to account for "hands free" activation or deactivation once an individual steps into the shower confinement. However, during the bench test of the prototype, an alternate HC –SR04 (Ultrasonic Sensor) and DHT 11 (Temperature and Humidity Sensor) were used in the initial proof of concept. The prime objective of the temperature and humidity sensor were to collect the real time relevant data regarding the temperature and humidity fluctuations occurring in the external environment and thereby adjusting the antithetical water discharge temperature. The temperature variation scenario was replicated in the classroom environment via an external stimulant such as the hairdryer. The data obtained are discussed in *Section IV: Results and Evaluation of data*. The SMART shower provides two major advantages during the initial implementation stage: (1) The temperature of the water discharge is adjusted in real-time and is not sustained on factual data that can have statistical outliers in each month (fluctuating temperatures during the summer or winter seasons). As discussed earlier, for additional safety a maximum water threshold can be set to $45^0$C or lower to avoid scalds. The temperature can also be adjusted to individual user preferences (such as *User A* always prefers warm shower irrespective of the season). (2) The ultrasonic sensor ensures that the water is conserved until the patron steps in proximity in the shower. The theory was tested in the laboratory setting with the defined near proximity range set to turn on was 1.5 feet and the deactivation control range set for 2.5 feet. A flowchart representing our proposed architecture is shown in Fig. 5.

**Fig.5 HERE**

As shown in the flowchart in Fig. 5 during the testing period, the temperature and humidity threshold was set to $23^0$C and 10% respectively. The proximity activation range was set to 1.5 feet and deactivation to 2.5 feet respectively.

The ultrasonic sensors can also be used to detect an “accidental fall” in the bathroom vicinity. Strategic placement of the sensors along the shower head pole can be critical in determining if the patron has fallen flat on the bathroom floor. Fig. 6 can be implemented in the proposed architecture to ensure accurate prediction of the accidental fall. If the ‘Ultrasonic sensor-1’ and ‘Ultrasonic sensor-2’ are not recording an individual’s presence but ‘Ultrasonic sensor-3’ is recording an obstacle; it can be easily verified as a slick fall. It can be further correlated with the occupancy time for an accurate extrapolation. A signal can be immediately sent for assistance to the caretaker.

**Fig. 6 HERE**

PAJ 7620 gesture recognition sensors can be used in conjunction with ultrasound sensor to monitor if a person is *“calling for help”*. A simple gesture such as moving an index finger right can signal that the person needs help whereas moving an index finger left could mean “*he/she is okay*”. The following code can be used to detect a right-hand movement which would turn on a red LED and print (“*Signal HELP*”) whereas a left hand would turn off the red LED and print (“*Patron is okay*”).

```
unit8_t data=0                // the Bank_0_Reg_0x43/0x44 should be read for hand gestures

paj7620ReadReg(0x43,1, &data)  // The data will take in the value depending on the  motion of

                               // the hand gestures

if (data = = GES_RIGHT_FLAG)     // If the gesture is movement of the right finger or hand

        digitalWire(4 , HIGH)      // Turns on red LED

        Serial.printIn(“Signal HELP”)

else (data = = GES_LEFT_FLAG)     // If the gesture is movement of the left finger or hand

        digitalWire(4 , Low)      // Turns off red LED

        Serial.printIn(“Patron is okay”)
```

Kim *et al.* (2018) developed a deep learning-based gesture recognition scheme (deepGesture) using motion sensors [25]. The model uses a new arm gesture recognition method based on gyroscope and accelerometer sensors that used deep convolution and recurrent neural network to automate learning of the sensor data. A similar deep learning model can extend to our prototype for accurate gesture recognition by our patron. This will allow more accurate sensing and be able to predict reliably;  hence reducing the chances of any false alarm.

CZN 15E, sound sensor can also be used to recognize if a patron is calling for help. CZN 15E is a binary sensor and if a patron in the shower space is repeatedly calling for “help”, it would display a series of binary 1 as opposed to 0 when there is no cry for help. The code below will display *‘1’* if the individual is calling and *‘0’* if the individual is quiet.

```
const int OUT_PIN =8                // The output is pin 8

void loop (){

Serial.printIn(digitalRead(OUT_PIN))
```

### B. *Cloud Server*

The IoT platform used for bench testing the prototype’s application is ThingSpeak™ . It is an open IoT platform with MATLAB analytics and hence is ideal to communicate and evaluate the sensor data and trigger response through actuators. The cloud server is primarily used as a relay for performing extensive data analysis and algorithm development as shown through a schematic in Fig. 7. IEEE 802.11n module was used to connect to the next layer in the architecture which will allow monitoring of the data in real-time. It will be able to process all the data in real-time collected through the hardware module. All this data can also be accessed and monitored through a dynamic web page, or the graphical mobile user interface proposed in Fig.2.

**Fig. 7 HERE**

The data are collected in the ThingSpeak channels from the Raspberry Pi connected sensors and help in the visualization of the data. The channels store all the data and can be viewed in real time. Each channel can include up to eight fields ( such as temperature, distance, humidity, etc.) For this bench testing, the channel settings were kept private but there are options where the channel can be made either public or shared with registered users. There are unique API keys that help to read or write data to a particular channel. The following code was compiled in the python script to connect and read the data from the temperature/ humidity and ultrasonic transducers.

```
key = 'HPUUWHG5GH8A768I' #Enter your Primary Channel WRITE API Key

READ_API_KEY='MUBAYFQE1GLGT631' #Enter your Observation Channel READ API Key

CHANNEL_ID='362929' #Enter your Observation Channel ID Number

entry_ID=0 #Sequence number of the receiving Observation Channel

sleep = 1 #How many seconds to sleep between posts to the channel
```

### C. *Algorithm Development*

The GPIO pins for the raspberry pi were defined in the python code as shown below. The sensors and actuators collected the corresponding relevant data and relayed it to the IoT platform, ThingSpeak$^{TM}$. The data over the temperature and humidity were monitored in real time though the platform's dynamic web page and are discussed in detail the following section: *Results and Evaluation of Data*.

```
#Set GPIO Pins

GPIO_TRIGGER = 18

GPIO_ECHO = 24

GPIO_LED = 16

GPIO_HOT = 26    // LED connected for HOT water discharge

GPIO_cold = 19  // LED connected for cold water discharge

GPIO_normal = 20 // LED connected for normal temperature water discharge

#set GPIO direction (IN / OUT)

GPIO.setup(GPIO_TRIGGER, GPIO.OUT)

GPIO.setup(GPIO_ECHO, GPIO.IN)

GPIO.setup(GPIO_LED,GPIO.OUT)

GPIO.setup(GPIO_HOT,GPIO.OUT)

GPIO.setup(GPIO_cold,GPIO.OUT)

GPIO.setup(GPIO_normal,GPIO.OUT)
```

The following python if-else code is written to measure the distance using an ultrasonic transducer ( to detect a patron’s presence) and activate the corresponding water temperature nozzle( depending on external temperature and humidity measurement) that is represented via LED lights for this application. If the distance is below 60 cm measured via the ultrasonic sensor, it would record the corresponding external temperature and if it is below 22 $^{0}$C , it would turn on the hot shower because it is colder outside, and the patron would enjoy a warmer shower.

```
# Python if-else code to activate shower nozzle depending on the temperature
if d<60:
    if temperature<22:
        print("Turn on hot shower")
        GPIO.output(GPIO_HOT,GPIO.HIGH)
        GPIO.output(GPIO_cold,GPIO.LOW)
        GPIO.output(GPIO_normal,GPIO.LOW)
        print("")
        print("")
        print("")
```

The initial design is a simple proof-of-concept and an example of the data visualization for distance measurement and time from the ultrasonic transducer on ThingSpeak is shown in Fig. 8. The visualization can be consumer friendly through a Java enabled mobile graphical user interface (GUI). Data privacy and customization can be protected through username and password or by choosing "*share channel view only with the following user*" from the sharing option of the primary channel on ThingSpeak. This increases consumer convenience as individualized water discharge temperature can be implemented inside the python code. Initially a single temperature and humidity sensor was connected in a series loop to the ultrasonic sensor via the Raspberry Pi 3 for the laboratory testing prototype. The system can be improved via strategic placement of multiple ultrasonic sensors along the varied length of the shower head as discussed in Fig. 6.

**Fig. 8 HERE**

A python code was used to ensure the maximum water discharge temperature to be less than $50^0$C as it can circumvent scalds from excessive high-water temperature. This added enabled features can ensure *safety and reliance* among patrons. The water consumption and the flow rate can also be controlled according to an individual's preference. As an optional feature, an additional code can be set to trigger warning through beaming of an LED to show that excessive water culvert through monitoring the water consumption. This eventually makes the proposed model *environmentally friendly*.

For an initial bench testing of the prototype, the safety and the environmentally friendly features were evaluated through a modest approach. The proximity real time measurement from the single ultrasonic sensor was synonymous with exposure time in shower confinement. *Scenario 1:* The data from the hot discharge of water flow can be combined with the shower occupancy time and can hence be used to define a potentially dangerous scenario in shower space. The smart shower system can identify if a patron is in the shower space with prolonged hot water discharge and can therefore be considered as a sign of potential danger and would result in deactivating the water flow.

*Scenario 2:* If a patron falls on the slippery bathroom floor with a heavy thud it can be easily monitored through CZN -15E microphone-based sound sensor. Data sampling can be performed at a fixed intensity and the amplitude can be monitored to record any sudden spike due to a loud thud for a fall. This can trigger to detect if ultrasound sensor-3 is recording an obstacle whereas simultaneously ultrasound sensor -1 and 2 is not detecting a patron. Image analysis of the obstacle using convolution neural network (CNN) can be used to improve the model prediction and avoid any false alarm.

*Scenario 3:* Apart from the ultrasound sensor PAJ 7620 gesture recognition sensors can be used in conjunction with ultrasound sensor to monitor if a person is *"calling for help"*. A simple gesture such as waving right can signal that the person has fallen and needs help . This could trigger a message and the information can be shared with registered users or caretakers via the ThingSpeak platform.

## IV. RESULTS AND EVALUATION OF DATA

The prototype to test the functionality of the sensors and their ability to connect to the ThingSpeak platform has been tested in a laboratory setting. The HC-SR04 and DHT 11 sensor was used to collect the relevant data which included: distance, temperature, and humidity in real-time. The data collected from the sensors were analyzed on the ThingSpeak platform through a compiled python script.

The architecture of the built prototype with the sensors connected to the Raspberry Pi and ethernet cable was used to connect to a primary computer to use the IoT platform is shown in Fig. 9. The cold, hot and normal temperature shower nozzle actuators were replaced by three LEDs during the development of this project. TABLE I shows the corresponding LED outputs correlated to defined task. The onset of the blue LED was the actuator response to activate the water discharge because the presence of a patron is detected using the ultrasound sensor -1.

Yellow LED was used to show that it was hot outside and therefore the colder water nozzle would be activated. An option can also be used where the patron chooses their preferred water temperature , for example $37^0$C, irrespective of the season or outside weather. This preference can be protected with individual user ID and operated with your smart phone few minutes prior to stepping into the shower space.

**Fig. 9 HERE**

**TABLE 1 HERE**

The model was tested for all the three ranges of temperature: hot, cold, and average temperature. The temperature was varied in the classroom environment by using a hair dryer near the DHT 11 sensor module. As shown in Fig. 9, the blue LED indicated that the shower space was occupied by an individual and similarly the green LED was the identification for the operation of the hot shower (since the experiment was conducted during winter). A *supplementary video S1* shows the actual experimentation and how the sensors were used to collect the data and relay to the IoT platform, ThingSpeak; the corresponding data analysis performed, and the response executed by the actuators (in this case the LEDs).

The data from the temperature and humidity sensor DHT-11 were taken in real time as a proof-of-concept that a connection was established between the Raspberry Pi, sensors and the IoT platform. Simultaneously, HC-SR04 was used to analyze an obstacle within proximity. The data received by the Raspberry Pi 3 was then relayed to ThingSpeak™ and a MATLAB data analytics of humidity, temperature and distance is shown as a function of time in Fig.10 (a)-(c).

**Fig. 10 (a) HERE**

**Fig. 10 (b) HERE**

**Fig. 10 (c) HERE**

The IEEE 802.11n Wi-Fi module was used by the Raspberry Pi 3 to connect to the IoT platform. The variation in Humidity (recorded in %), Temperature (recorded in $^{o}$ C) and distance (recorded in cm) is monitored through a primary computer by logging into ThingSpeak$^{TM}$ Channels. The obtained graphs in Fig. 10(a)-(c) verify that the real time data transmission occurred between sensors, Raspberry Pi 3 and the IoT platform. To validate the temperature fluctuation in classroom environment, the system was also tested by blowing a hot hair dryer near the temperature sensor for a few seconds and hence a temporary fluctuation in the temperature waveform were recorded in Fig. 10 (b) between 23 to 25 degrees Celsius. This indicated the switching to the cold shower mode as shown in Fig. 11 (b). A consumer-friendly graphical mobile interface is shown in Fig. 11(a) and (b) which was implemented through a simple python code to display the data analytics on the computer monitor after every 30 seconds' update.

**Fig. 11 (a) HERE**

**Fig. 11 (b) HERE**

The second stage of the project included using MATLAB data analytics to verify consumer *safety* features through data obtained via the sensors. Since the data were displayed as a function of time using MATLAB graphical interfacing, it was easier to track the shower occupancy period. A MATLAB algorithm with a condition based 'while loop' was used for the HC-SR04 ultrasonic sensor. Fig. 12 shows the output on the IoT platform screen. Fig. 12 depicts the test performance of the ultrasound sensor -1. If an individual is in the shower space , the GUI is going to print "*Shower space is occupied* " and when the patron leaves the following message is going to be printed "*Shower space is empty* ". An if-else loop can be used for the following condition: if ultrasound sensor -1 and 2 is reading “Shower space is empty ” but if ultrasound sensor-3 is reading “Shower space is occupied ”; a series of action as discussed in Scenario -1,2 and 3 can be executed. The date, period and occupancy status are clearly presented for the user’s convenience.

**Fig. 12 HERE**

The initial data is from a single source ultrasonic sensor but as discussed in Section III, simultaneous operation of three or more ultrasonic sensors placed at different altitudes of the pole of the shower will accommodate for additional safety features. A condition based ‘if-else’ algorithm from the three or more parallel operated ultrasonic sensors can accurately predict accidental fall.

## V. CONCLUSION AND FUTURE WORK

This paper presents the architecture prototype for a low-cost, energy efficient smart shower system that can be easily implemented to an existing system. The system focuses on three primary objectives : (1*) consumer convenience (2) environmentally friendly and (3) safety and reliance.* An initial proof-of concept of the system is tested to validate dynamic data transfer occurring between sensors/actuators, Raspberry Pi and the IoT platform. The temperature, humidity and ultrasonic sensors can allow a hands-free patron experience and convenience. The shower occupancy period and the amount of hot water consumed can be tracked and monitored for energy efficiency. For advanced consumer convenience further algorithms can be compiled as discussed in *Scenario-1, 2 and 3* and represent scope for future work in this field.

The real-time monitoring of temperature/humidity and the *occupancy of shower space* can be vital information for various aspects including conservation of water, consumer convenience and improving safety. The algorithms can be easily developed using python to accomplish a more complex task in the future. The advantage of the integration of MATLAB, python and the IoT platform allows remote monitoring and control. It hence represents a good prototype for commercial application after more complex functionalities are added to the initial model. Existing apps like Smart HQ can be used as a platform to operate and remotely control several smart appliances including our proposed low-cost, energy efficient and consumer safety smart shower system. Our future work incorporates the discussed features of *Section III* to improve consumer safety and reliance.

## Conflict of Interest

The authors have no conflict of interest.

## List of Table(s)

**Table I : The activation of the corresponding LED showing the execution of a specific signal using the sensor**

| **Command** | **LED color** | **Output** |
|---|---|---|
| Shower Space Occupied | Blue | ON |
| Shower Space Empty | Blue | OFF |
| Turn on cold shower | Yellow<br>Green / Red | ON<br>OFF |
| Turn on hot shower | Green<br>Yellow/ Red | ON<br>OFF |
| Turn on normal temperature shower | Red<br>Yellow/Green | ON<br>OFF |

## Figure Caption(s)

Fig.1: Front end schematic for '*consumer friendly'* advanced smart shower system.

Fig.2: Graphical Mobile User Interface for the proposed smart water monitoring system.

Fig. 3 : Raspberry Pi and the relevant sensors for the smart shower prototype

Fig.4 : Flowchart showing how headless Raspberry Pi can be connected to the IoT platform.

Fig.5 : Flowchart for proposed architecture for temperature and humidity measurement.

Fig. 6: The strategic placement of ultrasonic sensors on shower head pole to detect accidental fall in the shower space.

Fig. 7 : Schematic of how the data is collected from Raspberry Pi and processed at IoT platform ThingSpeak.

Fig. 8 : The Primary channel view of the distance-time measurement using an ultrasonic sensor as visualized on ThingSpeak.

Fig. 9 : Built in Architecture of SMART Shower model.

Fig. 10: The following output were obtained through MATLAB Analysis on ThingSpeak$^{TM}$.

(a) The graph shows the variation of Humidity (recorded in %) over real-time.

(b) The graph shows the variation of Temperature (recorded in $^{0}$C) over real-time

(c) The graph shows the variation of Distance (recorded in cm) over real time.

Fig. 11: Output of processed data by Raspberry Pi to the connected monitor screen.

(a)The reading on monitor when shower space is empty

(b)The reading on monitor when shower space is occupied.

Fig. 12 : Ultrasound Sensor-1 data on ThinkSpeak platform that can be interfaced to a dynamic webpage for ease of monitoring

## Alt Text for Figure(s)

Fig. 1 : Alt text : A layout of shower space confinement showing the placement of the microcontroller and ultrasonic sensor.

Fig. 2 : Alt text : A smart phone showing details about (a) maximum and minimum water temperature, (b) water used and the targeted water usage and (c) shower space occupancy time .

Fig. 3 : Alt text : Picture of Raspberry Pi 3, DHT 11, HC SR04 and CZN 15E

Fig.4 : Alt text : Flowchart showing how headless Raspberry Pi can be connected to the IoT platform

Fig. 5: Alt text: Flowchart for proposed architecture for temperature and humidity measurement

Fig. 6 : Alt text : Picture showing an individual lying on the floor and the placement of three ultrasonic sensors on a pole.

Fig. 7: Alt text : Picture showing a Raspberry Pi, along with ultrasonic sensor, temperature/ humidity sensor, sound sensor. An arrow is drawn from the Raspberry pi to a cloud labelled IoT platform ThingSpeak. Another arrow is drawn from the cloud to a flowchart followed by a MATLAB symbol

Fig. 8: Alt text : A chart labelled Primary channel showing distance( y-axis) and Date (x-axis).

Fig. 9 : Alt text : A breadboard wired with ultrasonic sensor, LEDs and temperature and humidity sensor. The blue and green LEDs are glowing.

Fig. 10(a): Alt text: A graph showing humidity (y-axis) and time (x-axis). It is a flat line at 15%.

Fig. 10 (b): Alt text: A graph showing temperature (y-axis) and time (x-axis). The value varies between 23 and 25 degrees Celsius

Fig. 10(c): Alt text : A graph shown distance( y-axis) and time (x-axis). The distance falls from 140 to 6 at 12:28 .

Fig. 11(a) :Alt text : A mobile screenshot showing distance 144, temperature 25, humidity 15.0, 200 OK and Shower room empty.

Fig. 11(b) :Alt text: A mobile screenshot showing distance 16, temperature 23, humidity 15.0, 200 Ok, Shower Turned on, Temperature =23, Humidity =15% , Turn on cold shower.

Fig. 12: Alt text : A graph of distance( y-axis) and time( x-axis). “The person steps away from shower space” is marked at 12:12 (30 seconds). Their a section labelled output which shows “Shower space occupied” , Distance =8 (29-Nov-2017) and Shower space empty , Distance=74.

## List of Figure(s)

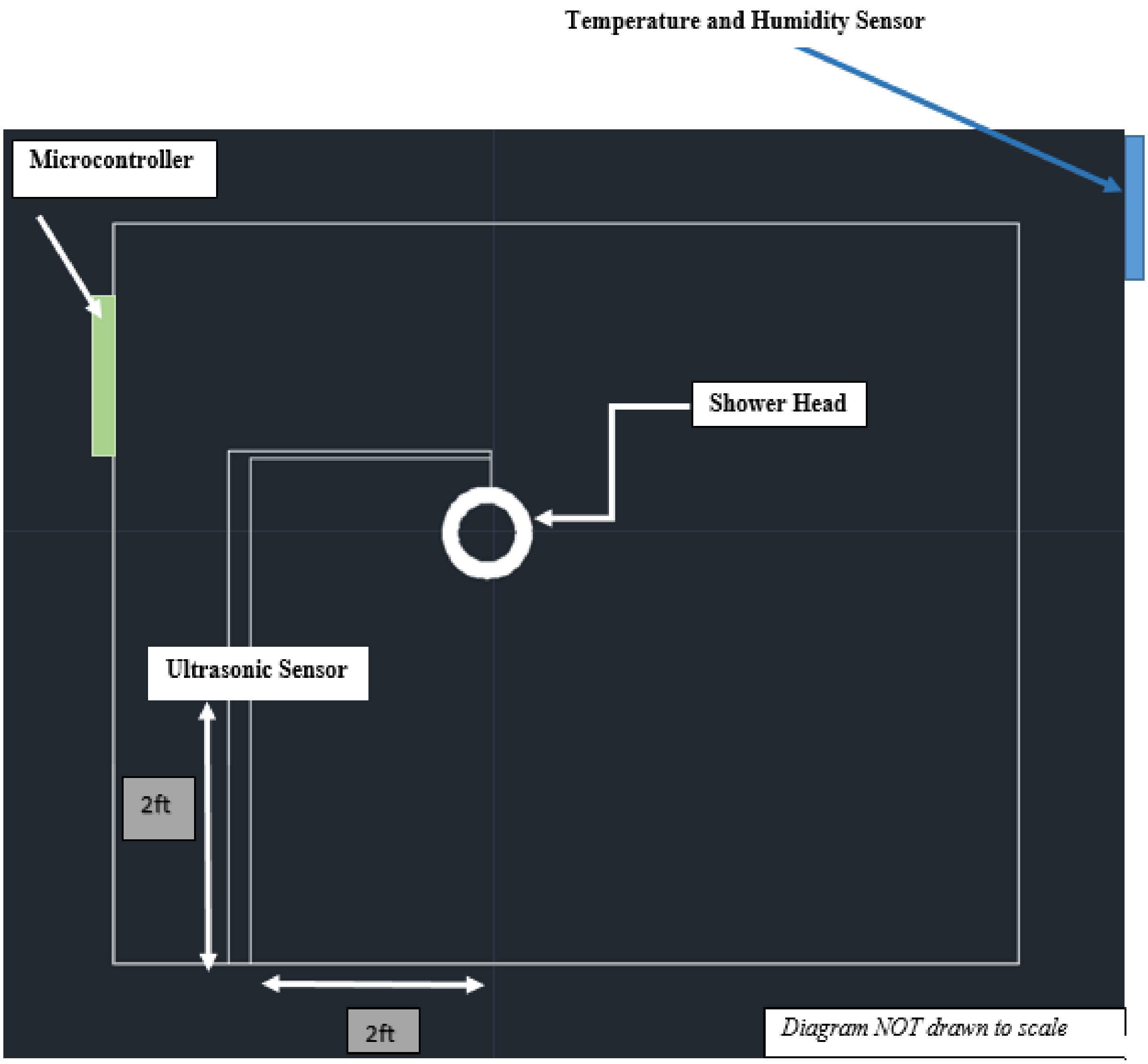

**Fig.1**

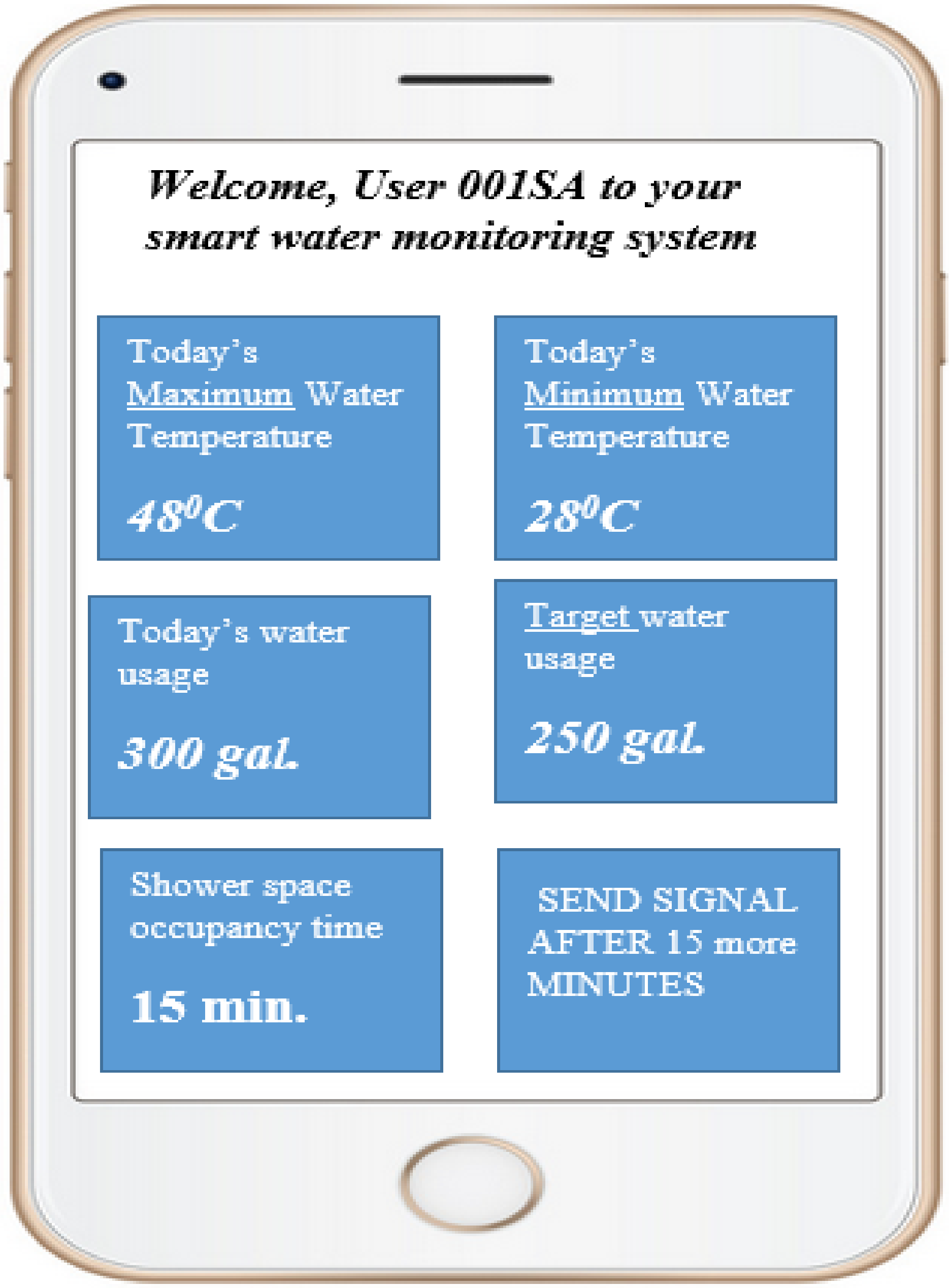
Welcome, User 001SA to your smart water monitoring system
Today's Maximum Water Temperature
48⁰C
Today's Minimum Water Temperature
28⁰C
Today's water usage
300 gal.
Target water usage
250 gal.
Shower space occupancy time
15 min.
SEND SIGNAL AFTER 15 more MINUTES

**Fig. 2**

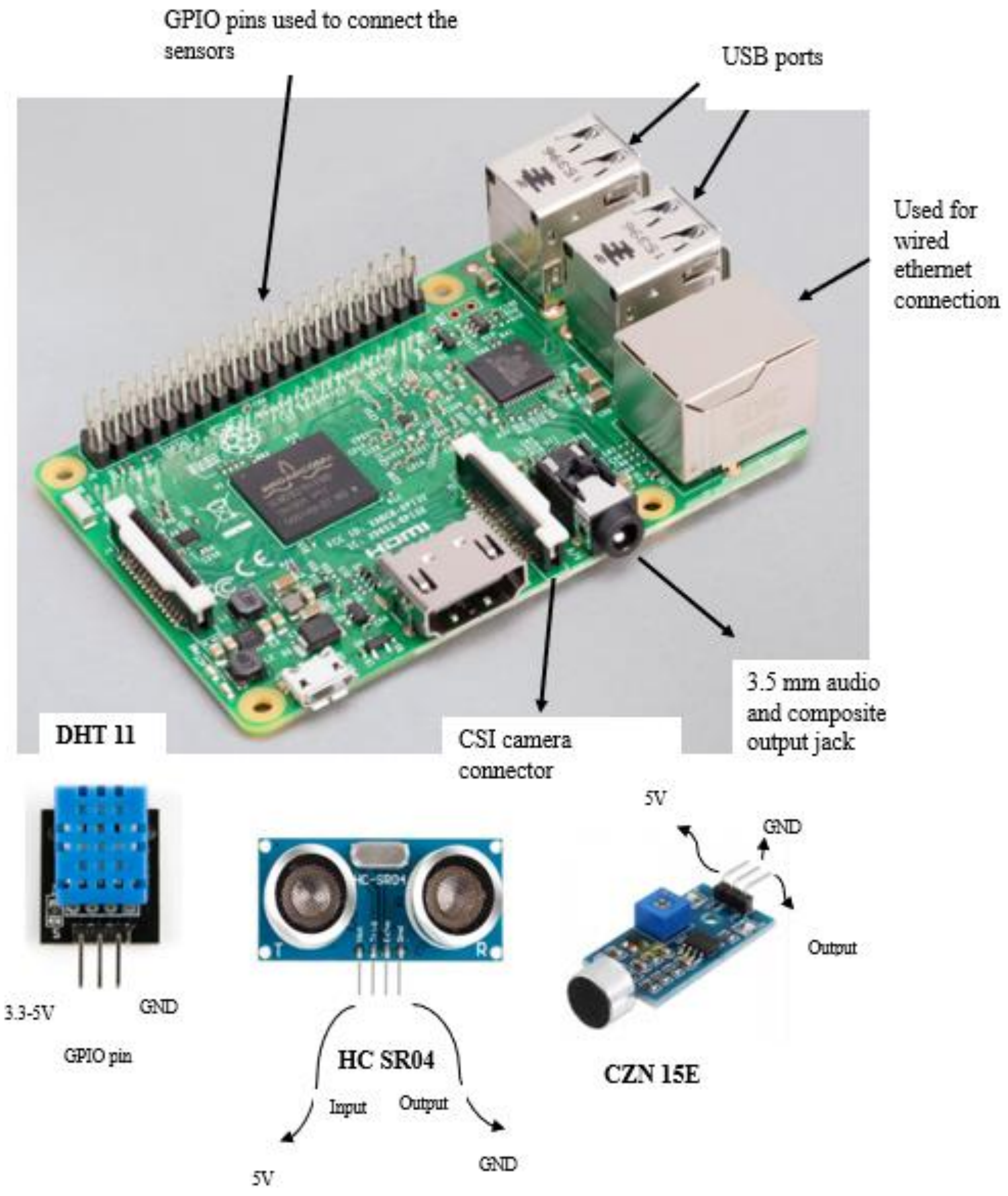

**Fig. 3**

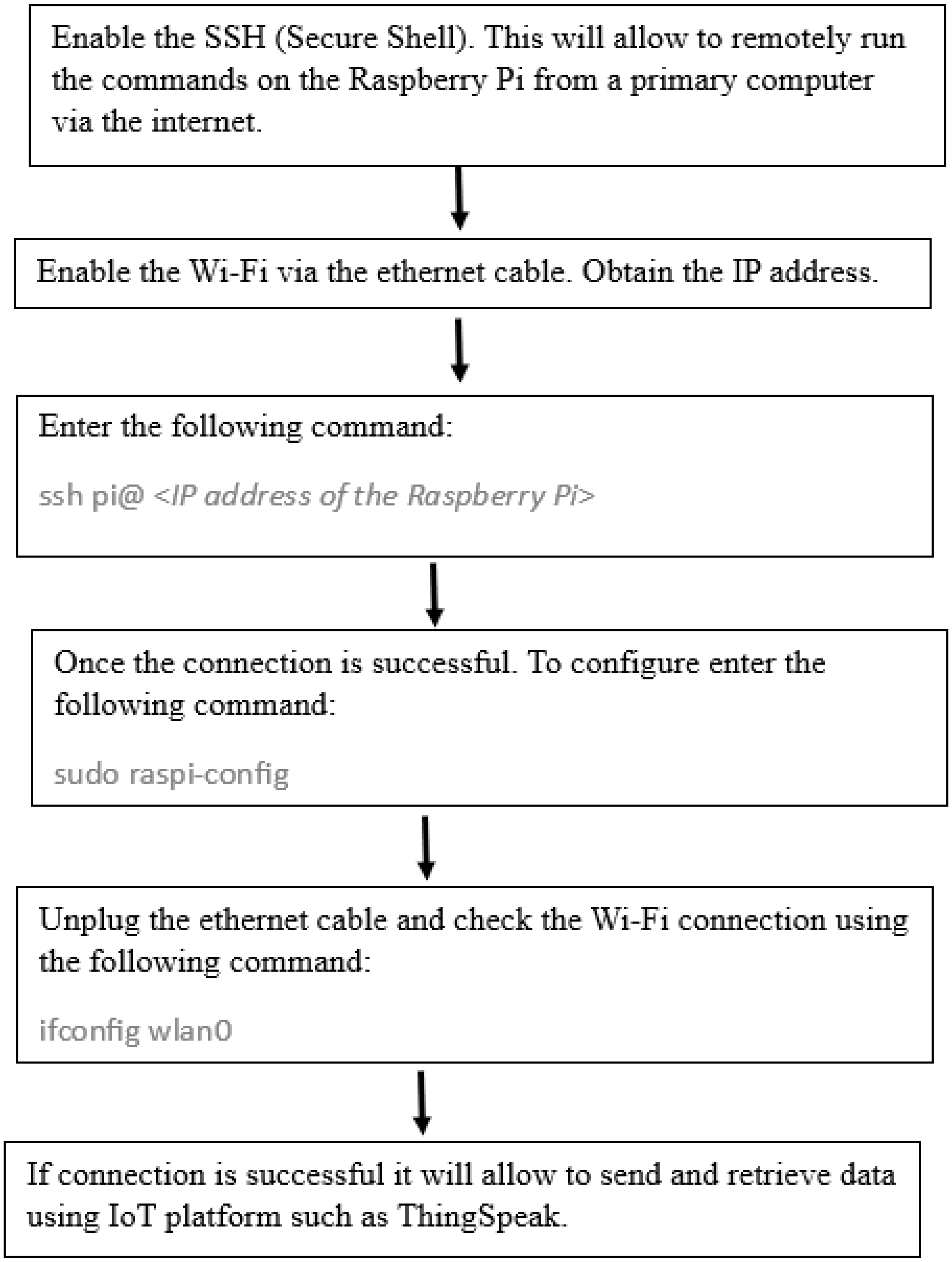

Enable the SSH (Secure Shell). This will allow to remotely run the commands on the Raspberry Pi from a primary computer via the internet.
Enable the Wi-Fi via the ethernet cable. Obtain the IP address.
Enter the following command:
ssh pi@ <IP address of the Raspberry Pi>
Once the connection is successful. To configure enter the following command:
sudo raspi-config
Unplug the ethernet cable and check the Wi-Fi connection using the following command:
ifconfig wlan0
If connection is successful it will allow to send and retrieve data using IoT platform such as ThingSpeak.

**Fig. 4**

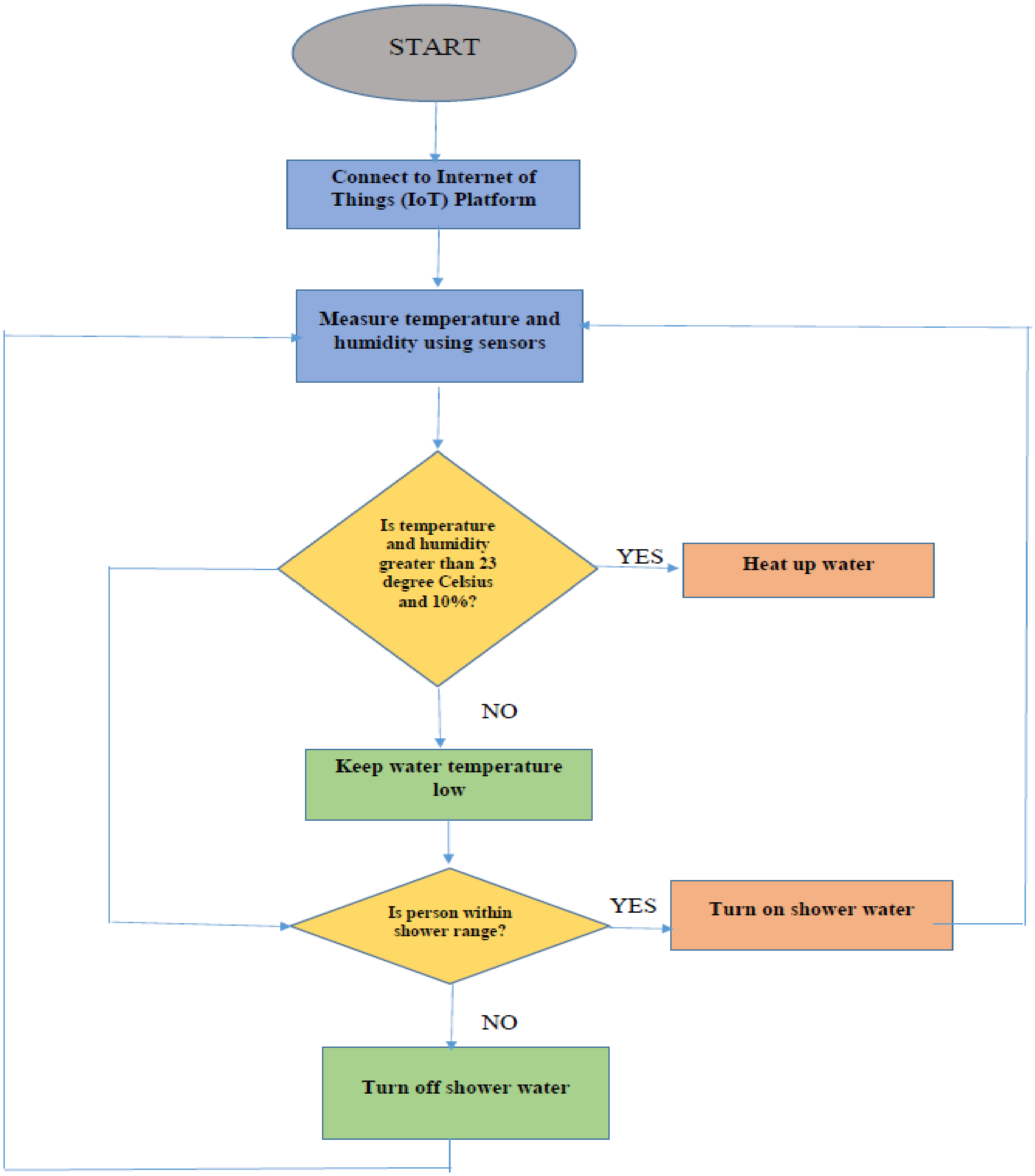
START
Connect to Internet of Things (IoT) Platform
Measure temperature and humidity using sensors
Is temperature and humidity greater than 23 degree Celsius and 10%?
YES
Heat up water
NO
Keep water temperature low
Is person within shower range?
YES
Turn on shower water
NO
Turn off shower water

**Fig. 5**

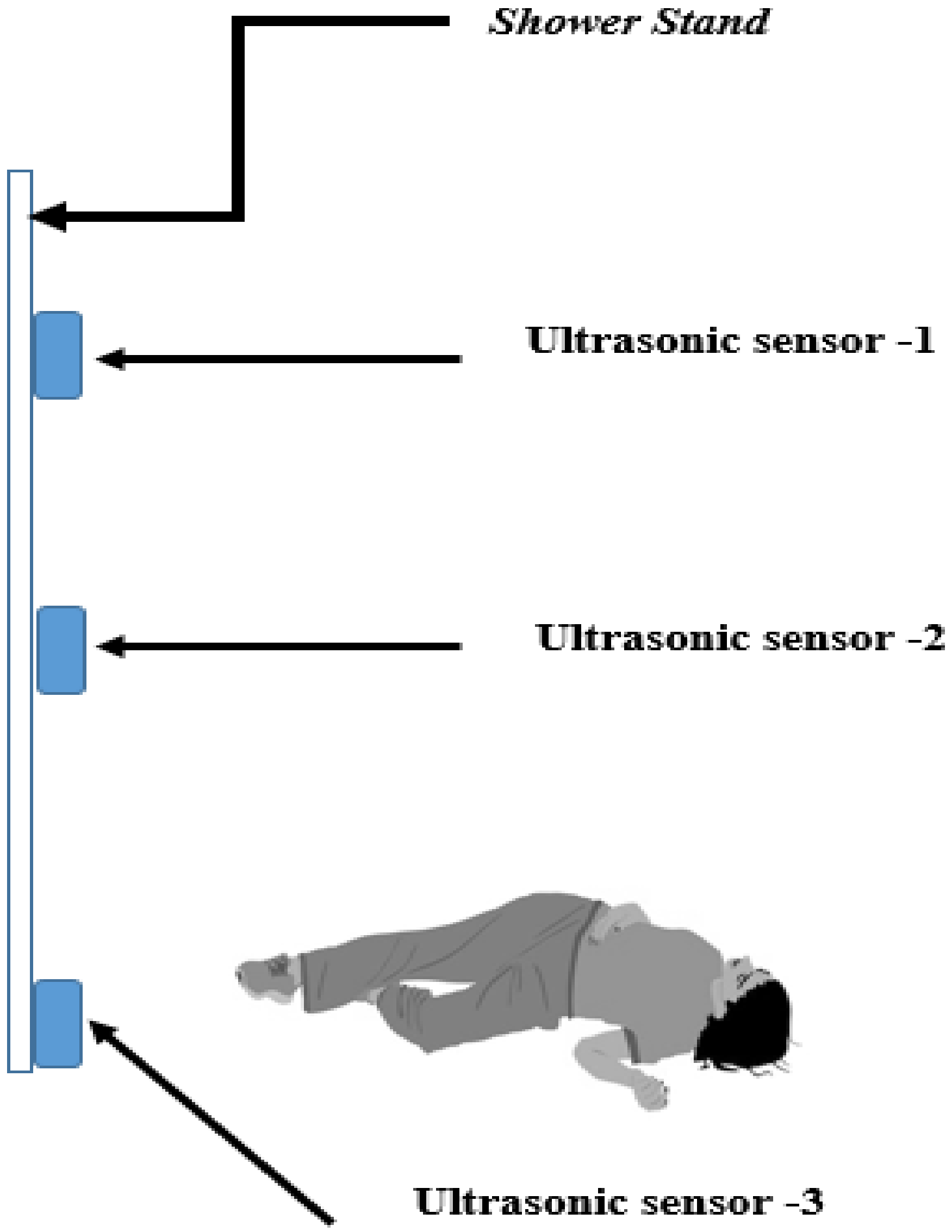
Shower Stand
Ultrasonic sensor -1
Ultrasonic sensor -2
Ultrasonic sensor -3

**Fig. 6**

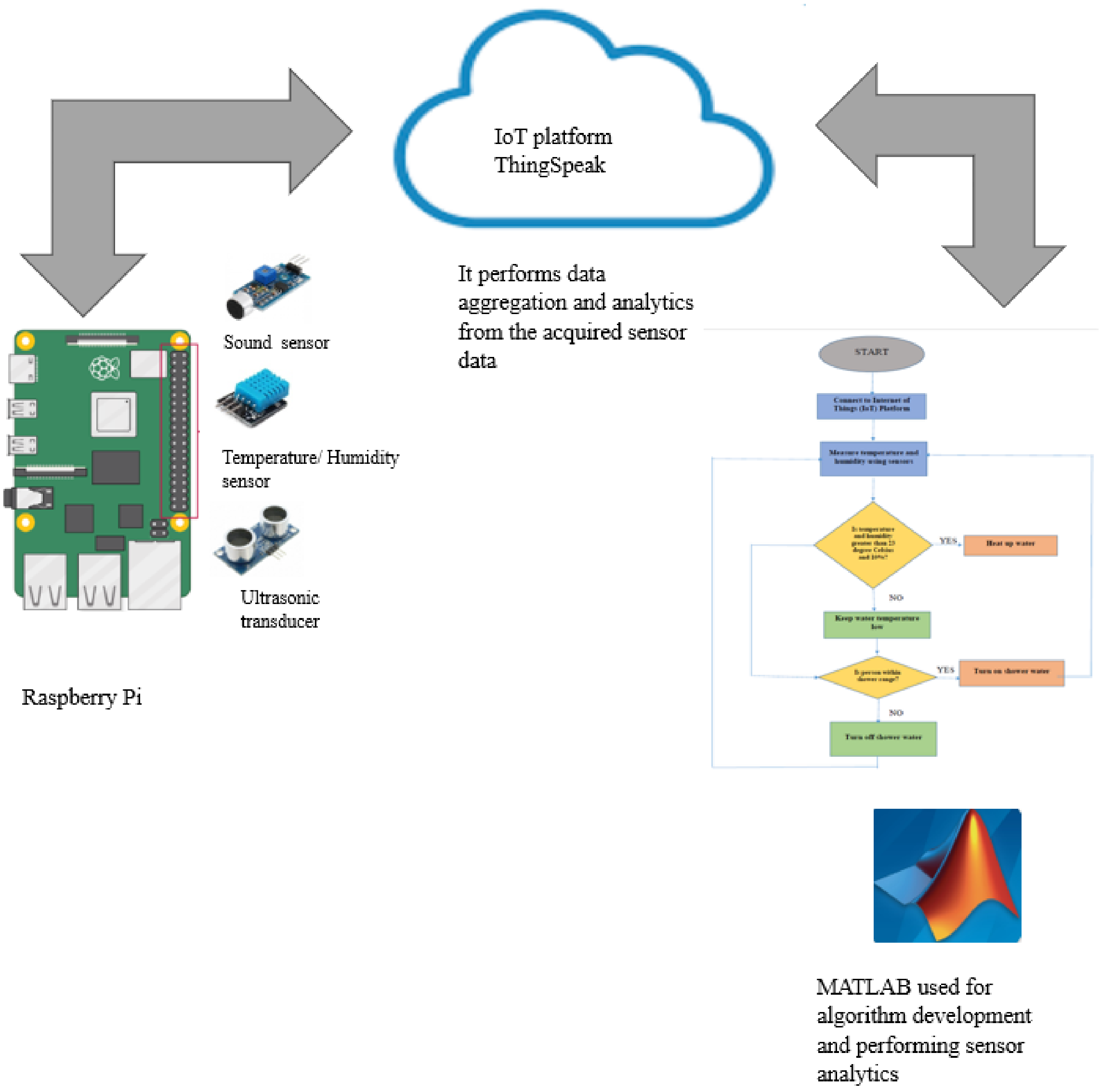
IoT platform
ThingSpeak
It performs data aggregation and analytics from the acquired sensor data
Sound sensor
Temperature/ Humidity sensor
Ultrasonic transducer
Raspberry Pi
START
YES
Heat up water
NO
YES
NO
MATLAB used for algorithm development and performing sensor analytics

**Fig. 7**

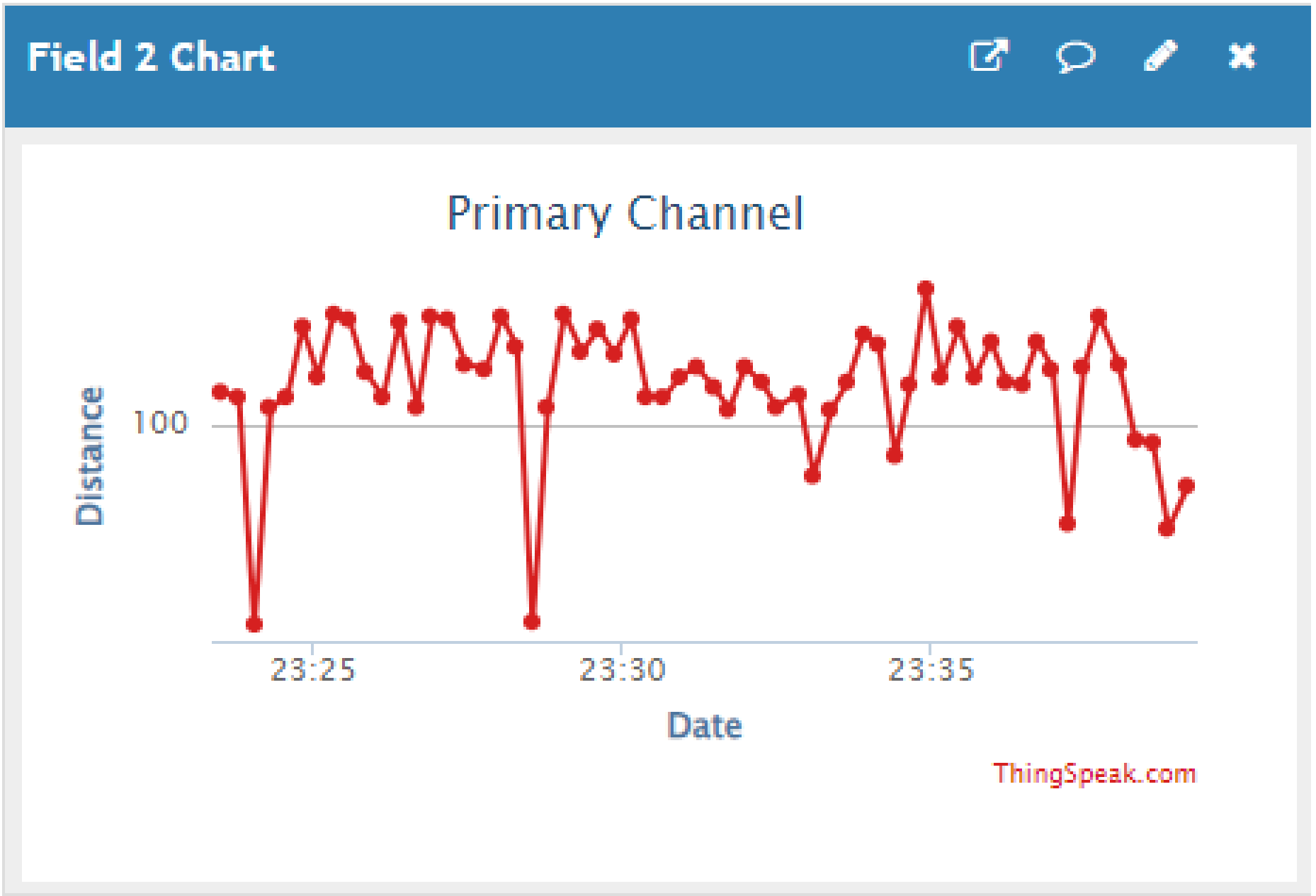

**Fig. 8**

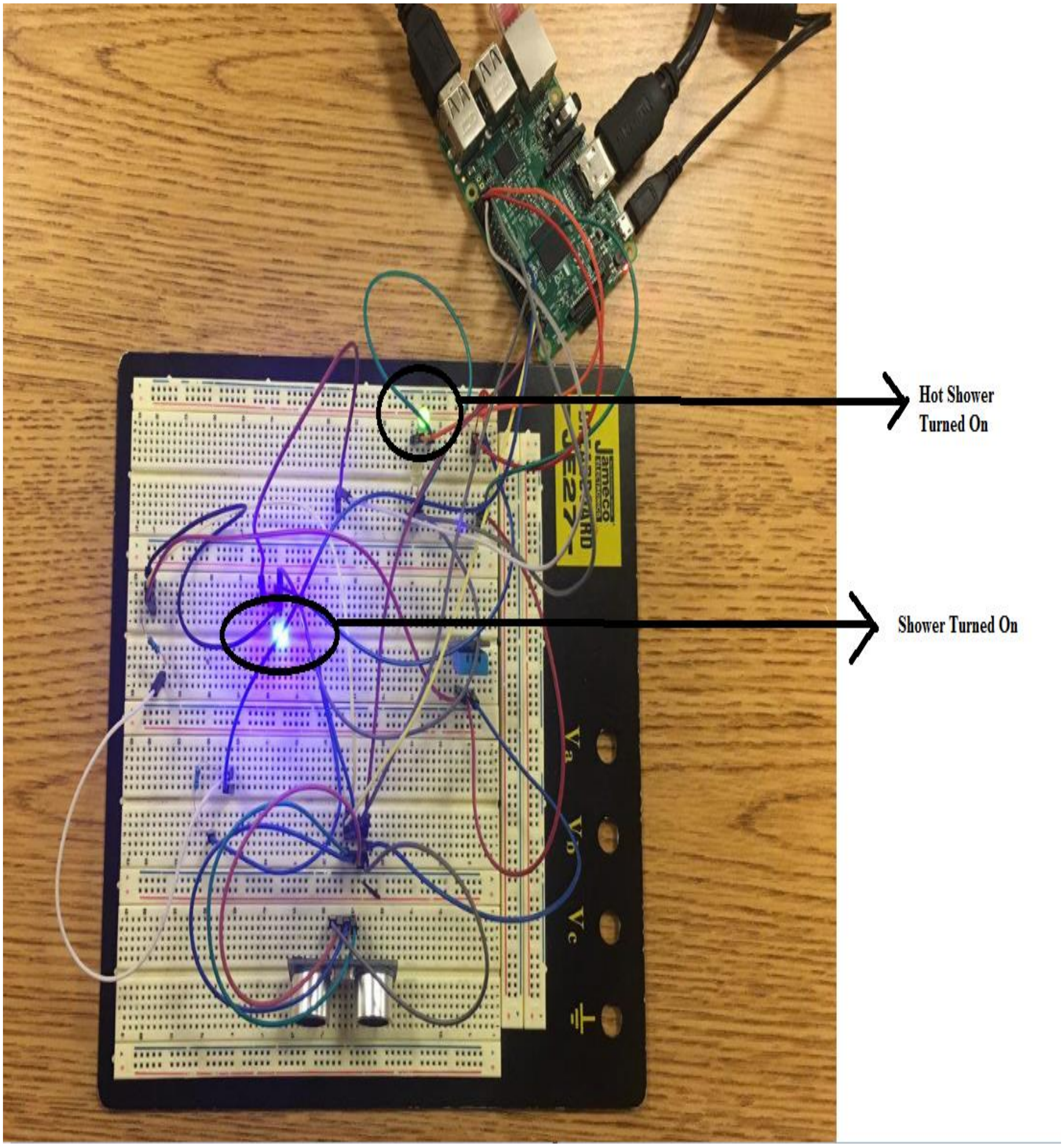

**Fig. 9**

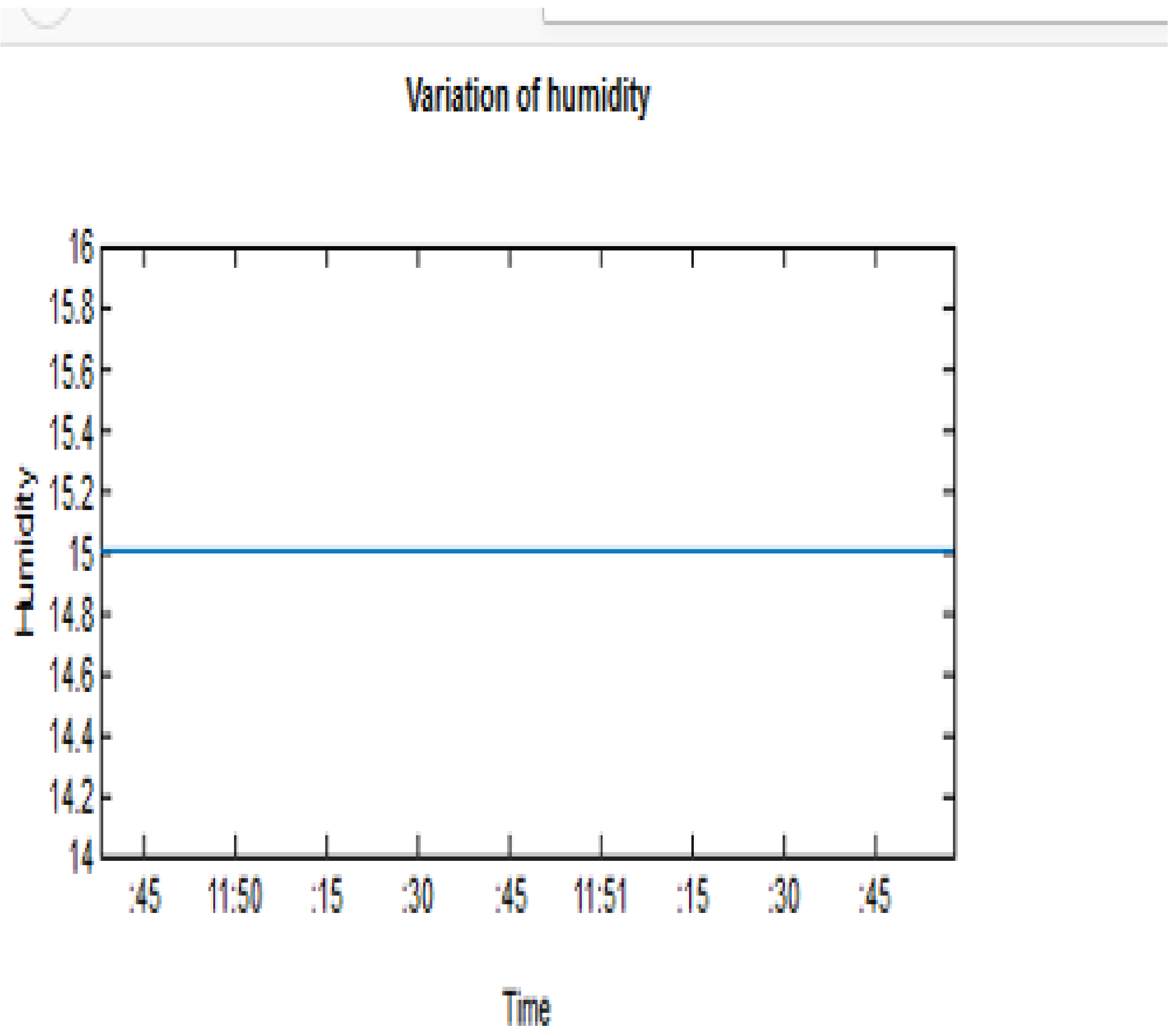

**Fig. 10 (a)**

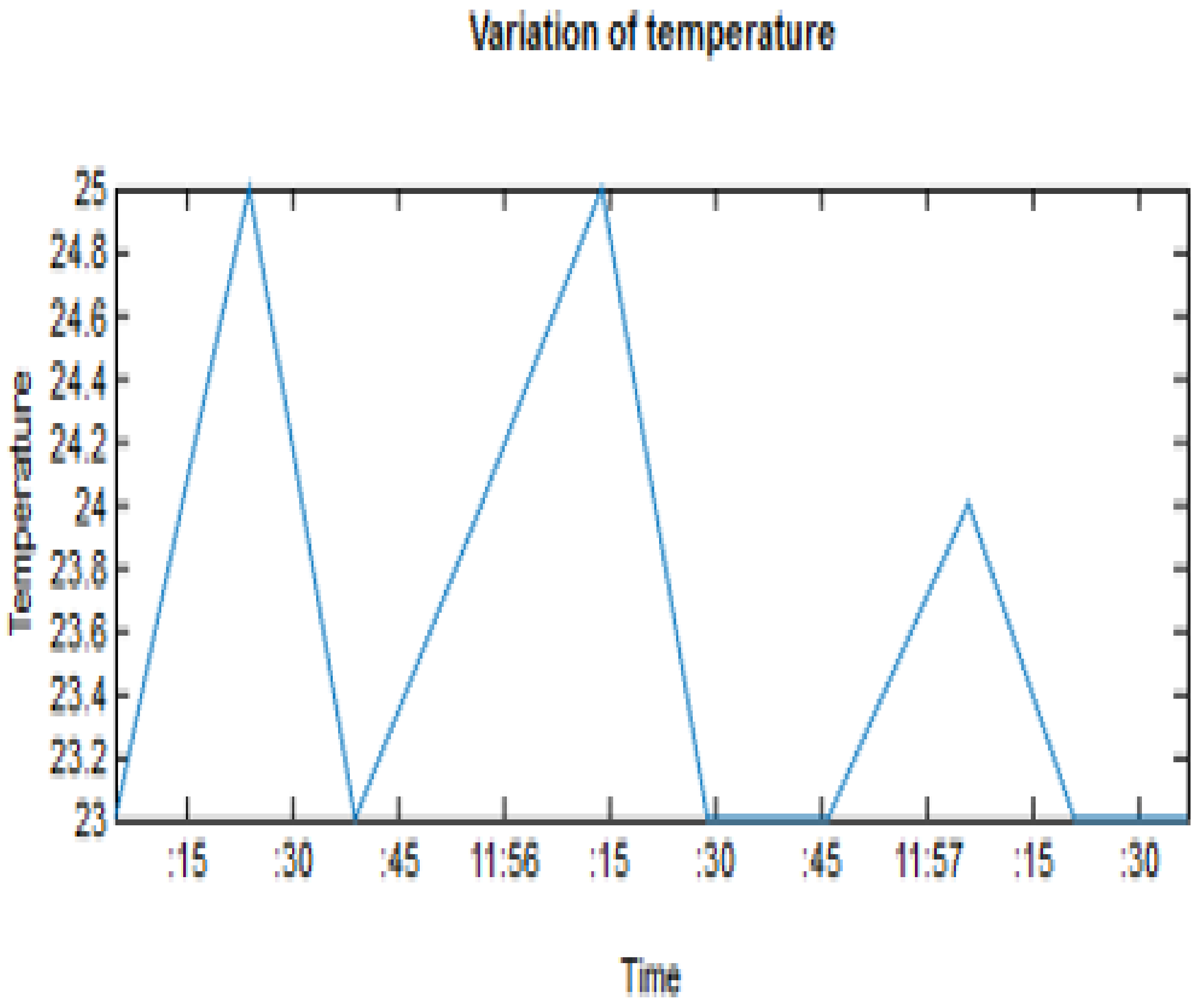

**Fig. 10 (b)**

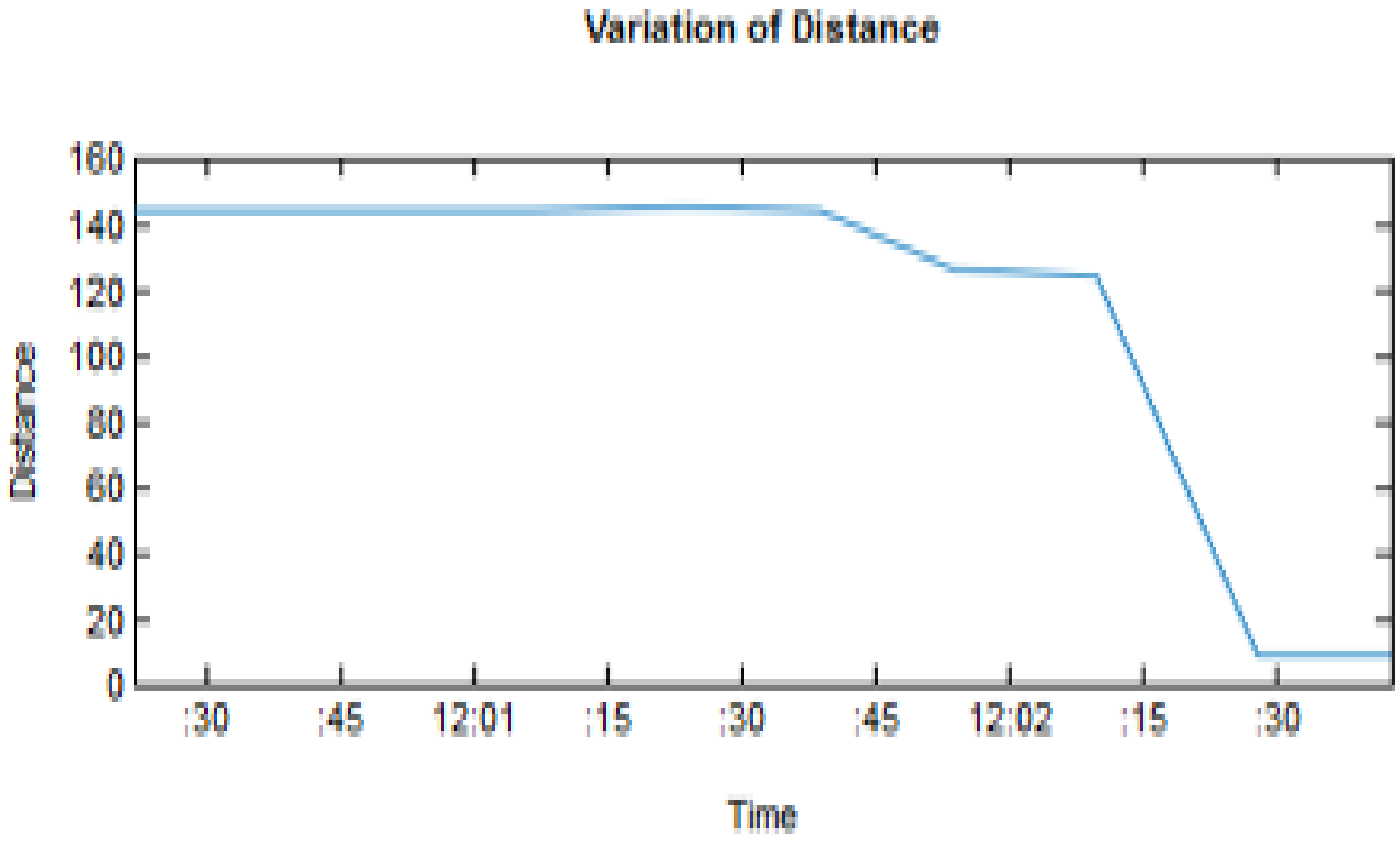

**Fig. 10 (c)**

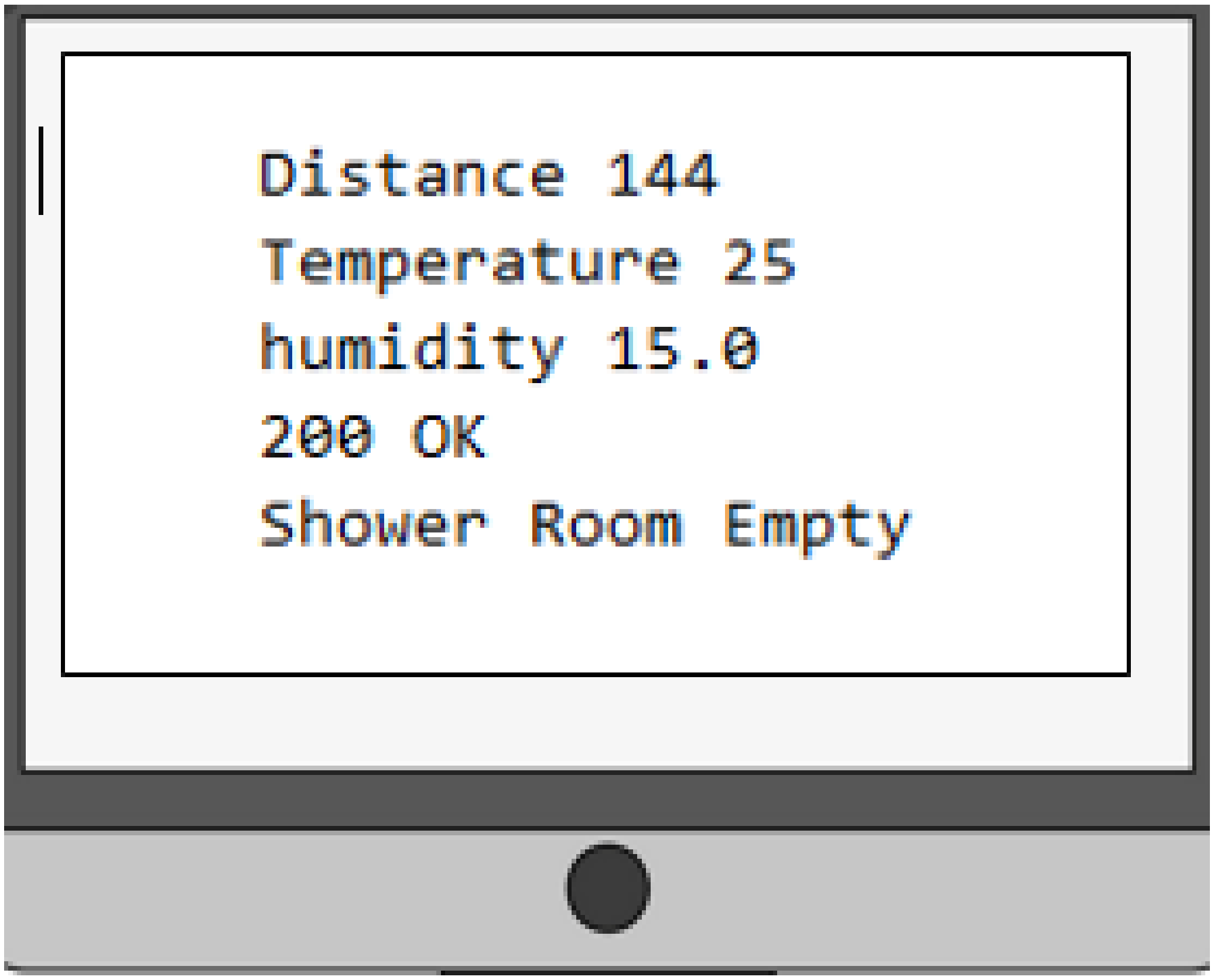
Distance 144
Temperature 25
humidity 15.0
200 OK
Shower Room Empty

**Fig. 11 (a)**

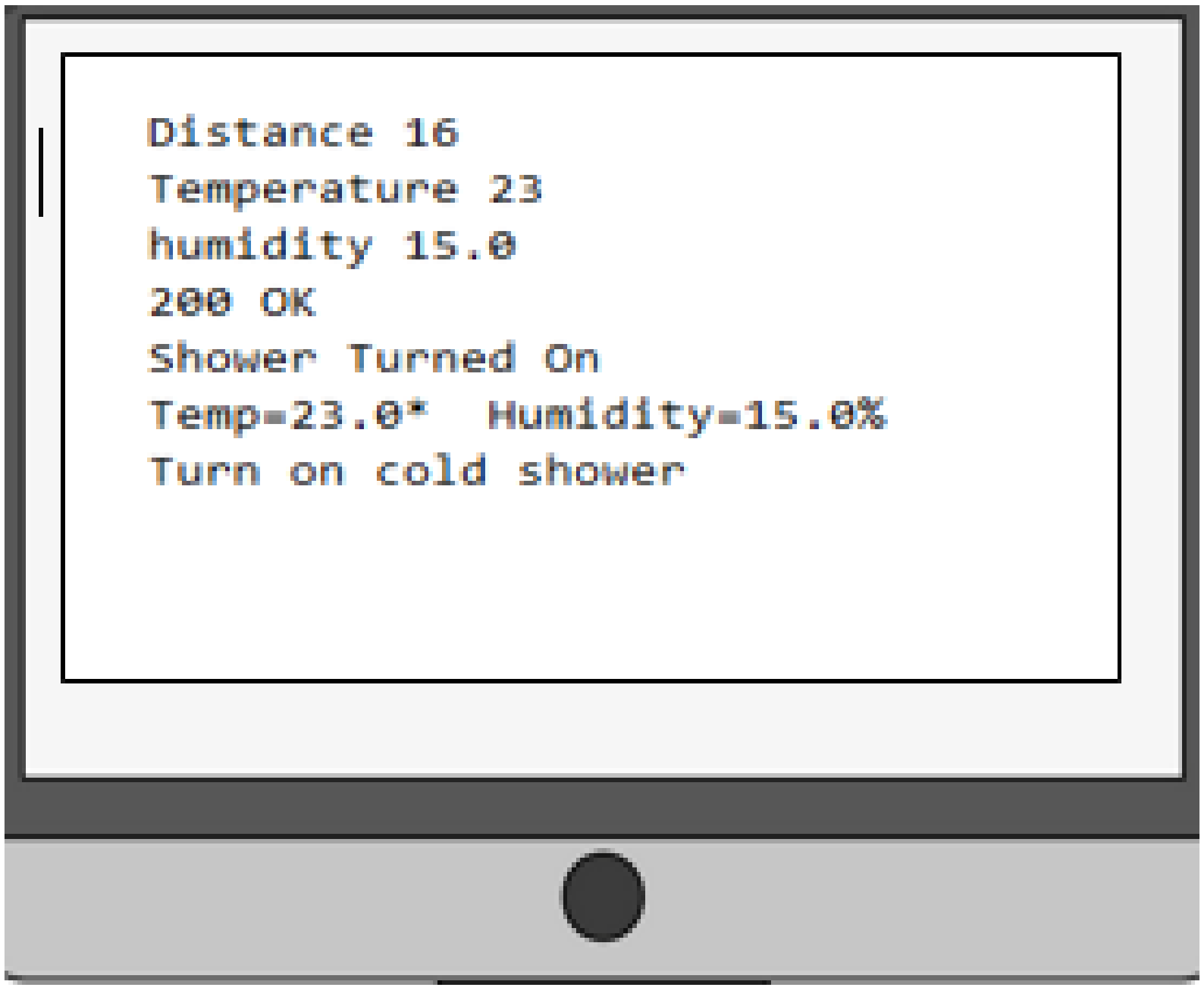

**Fig. 11 (b)**

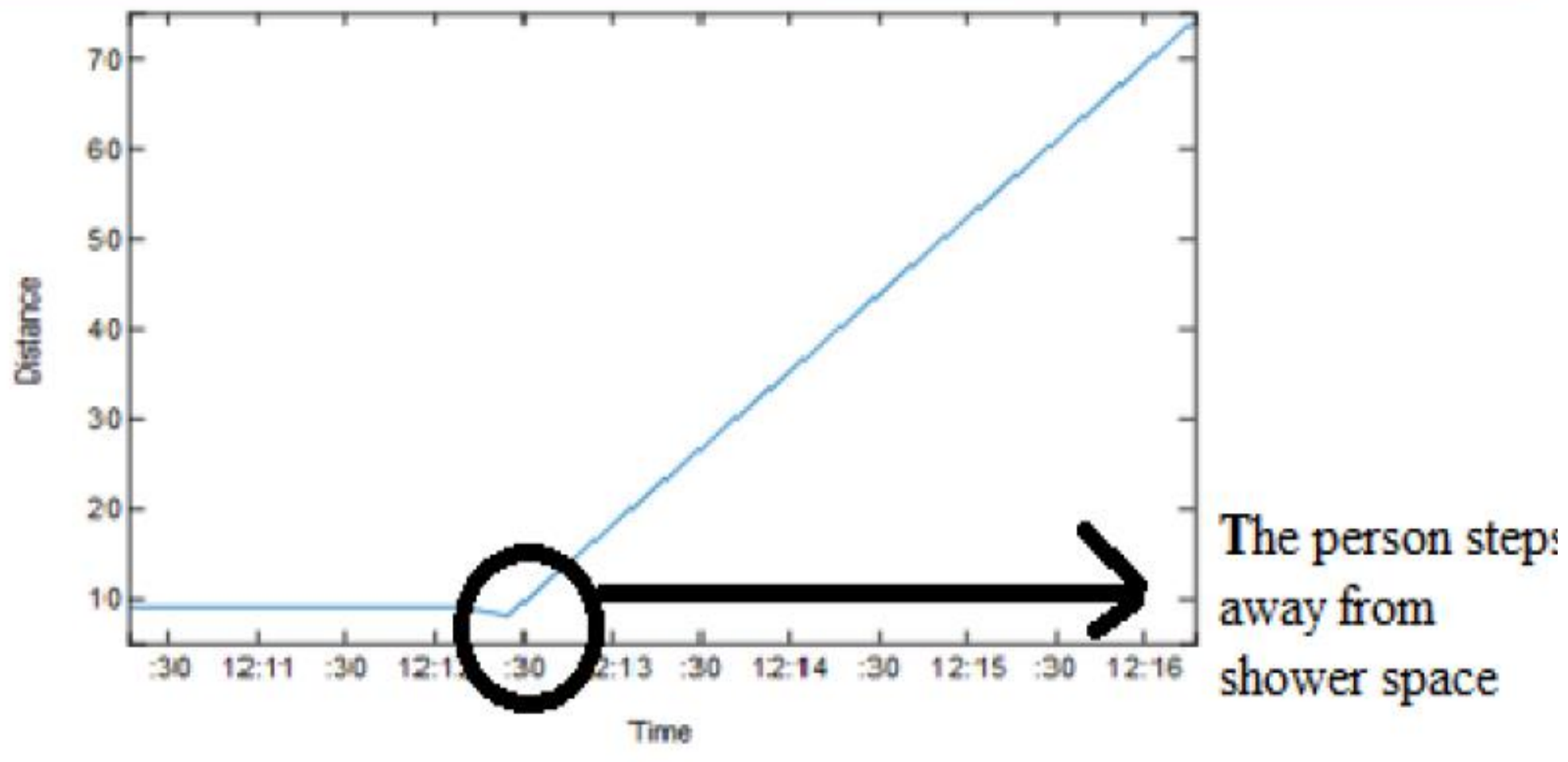

**Output**

```
Shower space occupied
Distance=
     8

   29-Nov-2017 05:12:25


Shower space empty
Distance=
    74
```

**Fig. 12**